\documentclass[aps, pra, twocolumn, superscriptaddress, english, floatfix]{revtex4}
\usepackage{graphicx}
\usepackage{cancel}
\usepackage{enumerate}
\usepackage{hyperref}
\usepackage{textcomp}
\usepackage{amsmath}
\usepackage{amssymb}
\usepackage{pgf}
\usepackage{soul}
\usepackage{lipsum}
\usepackage{comment}

\usepackage[english]{babel}
\usepackage{bm}
\usepackage{tikz}
\usetikzlibrary{shapes.geometric}
\usetikzlibrary{decorations.markings}
\usetikzlibrary{decorations.pathmorphing,patterns}
\usetikzlibrary{decorations}
\usetikzlibrary{calc}
\usetikzlibrary{intersections}
\usetikzlibrary{arrows}
\usetikzlibrary{matrix}
\usetikzlibrary{positioning}
\newcommand{\bbeta}{\boldsymbol{\beta}}
\newcommand{\bgamma}{\boldsymbol{\gamma}}
\newcommand{\bsigma}{\boldsymbol{\sigma}}
\newcommand{\bxi}{\boldsymbol{\xi}}

\newcommand{\nep}{\mathrm{e}}

\newcommand{\upket}{|\!\!\uparrow\rangle}
\newcommand{\downket}{|\!\!\downarrow\rangle}

\newcommand{\dQA}{\mathrm{\scriptscriptstyle dQA}}
\newcommand{\ncost}{\mathrm{n_{\scriptscriptstyle c}}}
\newcommand{\ans}{\mathrm{\scriptscriptstyle Ansatz}}

\newcommand{\target}{\mathrm{\scriptscriptstyle targ}}

\newcommand{\PauliSigma}{\hat{\sigma}}

\newcommand{\Ho}{\hat{H}}

\newcommand{\Hz}{\widehat{H}_z}
\newcommand{\Hx}{\widehat{H}_x}

\newcommand{\Ptrot}{\mathrm{P}}

\graphicspath{{./}{./Fig/}}

\begin{document}

\title{Quantum Approximate Optimization Algorithm applied to the binary perceptron}
\author{Pietro Torta}
\affiliation{SISSA, Via Bonomea 265, I-34136 Trieste, Italy}
\author{Glen B. Mbeng}
\affiliation{Universit\"at Innsbruck, Technikerstra{\ss}e 21 a, A-6020 Innsbruck, Austria}
\author{Carlo Baldassi}
\affiliation{Department of Computing Sciences, Bocconi University, 20136 Milan, Italy}
\author{Riccardo Zecchina}
\affiliation{Department of Computing Sciences, Bocconi University, 20136 Milan, Italy}
\author{Giuseppe E. Santoro}
\affiliation{SISSA, Via Bonomea 265, I-34136 Trieste, Italy}
\affiliation{International Centre for Theoretical Physics (ICTP), P.O.Box 586, I-34014 Trieste, Italy}
\affiliation{CNR-IOM, Consiglio Nazionale delle Ricerche - Istituto Officina dei Materiali, c/o SISSA Via Bonomea 265, 34136 Trieste, Italy}

\begin{abstract}
We apply digitized Quantum Annealing (QA) and Quantum Approximate Optimization Algorithm (QAOA)  
to a paradigmatic task of supervised learning in artificial neural networks: the optimization of 
synaptic weights for the binary perceptron. 
At variance with the usual QAOA applications to MaxCut, or to quantum spin-chains ground state preparation, 
the classical Hamiltonian 
is characterized by highly non-local multi-spin interactions. 
Yet, we provide evidence for the existence of optimal \emph{smooth} solutions for the
QAOA parameters, which are \emph{transferable} among typical instances of the same problem, and we prove numerically an enhanced performance of QAOA over traditional QA.
We also investigate on the role of the QAOA optimization landscape geometry in this problem, showing that
the detrimental effect of a gap-closing transition encountered in QA is also negatively affecting the performance of our implementation of QAOA.
\end{abstract}

\maketitle

\section{Introduction}


Quantum optimization is a very active branch of quantum computation~\cite{Nielsen_Chuang:book},  
which started with Quantum Annealing (QA)~\cite{Finnila_CPL94, Kadowaki_PRE98, Brooke_SCI99, Santoro_SCI02, Santoro_JPA06} and Adiabatic Quantum Computation (AQC) \cite{Farhi_SCI01, Albash_RMP18}, by now well established and implemented in analogue dedicated hardware~\cite{Johnson_Nat11}.

New research directions are currently pursuing the design of parameterized quantum circuits and Variational Quantum Algorithms~\cite{VQA_review}.
Among the early proposals in this class are the Variational Quantum Eigensolver (VQE)~\cite{VQE} and the Quantum Approximate Optimization Algorithm (QAOA)~\cite{Farhi_arXiv2014}, hybrid quantum-classical variational optimization schemes~\cite{McClean2016} designed for quantum ground state preparation, and classical combinatorial optimization, respectively. 
In particular, QAOA has rapidly gained popularity, with new theoretical understanding~\cite{Lloyd_arXiv2018, morales_arxiv2019, Mbeng_arXiv2019, brady2021behavior, Zhou_PRX2020}
and successful implementations on experimental platforms~\cite{Pagano_arXiv2019}. 

In a broader perspective, various Variational Quantum Algorithms have been proposed and applied beyond classical combinatorial optimization~\cite{SciPost_quantumGS, Carrasquilla},
conceptually generalizing QAOA~\cite{QAlternatingOA} or VQE~\cite{layerVQE}, or focusing on realistic quantum hardware setups~\cite{farhi2017quantum, VQE_hardware_efficient}. 
These algorithms constitute a rich playground for short-term implementations on existing and near-future noisy intermediate-scale quantum devices~\cite{Preskill_Quantum2018}, aiming at some type of quantum speedup~\cite{q_speedup}.

The quantum Hamiltonian of a physical system typically consists of a sum of local terms, each involving only a small subset of quantum degrees of freedom. 
On the contrary, in order to apply QA or QAOA to a classical optimization problem, an {\em embedding} into a quantum setup~\cite{Lucas2014} is first required. 
In this second case, the standard strategy is to start from a classical cost (or energy) function $E(\sigma_1,\dots,\sigma_N)$ depending on $N$ binary variables $\sigma_j=\pm 1$, and to map them onto quantum spin-$1/2$ Pauli operators $\hat{\sigma}^z_j$, hence regarding the initial cost function as a quantum Hamiltonian which is \emph{diagonal}, by construction, in the standard computational basis of quantum computation~\cite{Nielsen_Chuang:book}:
\begin{equation}\label{eq:embedding}
   E(\sigma_1,\dots,\sigma_N) \to \Ho_z(\PauliSigma_1^z,\dots,\PauliSigma_N^z) \;.
\end{equation}
Nevertheless, in contrast with the Hamiltonian of physical quantum systems, the embedding Hamiltonian $\Ho_z$ can be arbitrarily difficult to implement, possibly including non-local multi-spin interactions, thus proving intrinsically hard to simulate exactly~\cite{Lloyd_qsimulators} on a quantum device.

Remarkably, some specific combinatorial optimization problems admit a natural embedding into a $2$-local Ising spin-glass Hamiltonian, such as MaxCut on random graphs~\cite{Farhi_arXiv2014}, with few other model-specific applications~\cite{Wauters_PRA2020}.
As a matter of fact, many interesting classical optimization problems do not admit such a simple reformulation in terms of a $2$-local (or $k$-local, for any $k\ll N$) Hamiltonian.

An interesting example --- of paramount importance in machine learning--- is offered by the training process required in supervised learning for artificial neural networks (ANNs): this is naturally formulated as a minimization problem of a suitable cost function~\cite{sup_learning_review}, which is, however, non-local in terms of its variables (network weights) due to non-linear activation functions. 
An intriguing question is to explore {\em if} and {\em how} the laws of quantum mechanics might allow for particularly efficient algorithms to train ANNs, while potentially offering some deeper theoretical understanding of their effectiveness in classification tasks. 
From a more general perspective, recent years have witnessed the rise of an emerging field known as quantum machine learning~\cite{QML} and different proposals have been put forward to perform machine learning tasks by developing quantum versions of neural networks: interestingly enough, parameterized quantum circuits themselves can be regarded as alternative machine learning models~\cite{PQC_ML}.

In this work, we aim at investigating the potential applicability of Variational Quantum Algorithms such as QAOA in the realm of hard non-convex classical optimization problems, yielding highly non-local Hamiltonians, beyond the usual $2$-local models.
As a working example, we focus on the problem of learning random patterns in a single-layer neural network with binary weights, the so-called binary perceptron. 
Our work stems from the results obtained in Ref.~\cite{BaldassiPNAS2018}, where the authors provided analytic and simulation evidence of exponential speed-up of Quantum Annealing vs. classical Simulated Annealing for the training process of the binary perceptron. The exponential speedup arises from the geometric structure of the solution space of the problem: the presence of rare and yet dense regions of solutions allows Quantum Annealing to converge efficiently despite the presence of an exponential number of local minima. This property appears to be present even in more complex, highly overparameterized neural networks such as the so-called deep neural networks \cite{baldassi2021learning}. Quantum algorithms could thus be highly efficient also for this kind of models, which define the state of the art in contemporary machine learning. Here, we focus on a digitized version of Quantum Annealing and QAOA: in particular, we provide numerical evidence on how QAOA, by efficiently exploiting optimized quantum fluctuations among classical states, systematically outperforms standard QA.
As in Ref.~\cite{BaldassiPNAS2018}, these results are expected to generalize for more complex ANN architectures.

Moreover, we show the emergence of \emph{smooth} optimal QAOA parameters~\cite{Zhou_PRX2020,Mbeng_arXiv2019,Mbeng_optimal-dQA_arXiv2019}, which seem to be independent of the details of the training problem. 
This finding allows us to develop an effective heuristic procedure to speed up the convergence of QAOA, in a similar fashion to previous results for $2$-local models.
In fact, one of the most promising research lines in Variational Quantum Algorithms deals with reducing or removing the need for a classical optimization loop~\cite{streif2019training}, by leveraging on \emph{concentration effects} in the energy landscape for different problem instances of the same class~\cite{brandao2018, farhi2020quantum, galda2021transferability}, or for different sizes $N$ of the same instance~\cite{QAOA_concentration}. In our work we give numerical evidence of similar findings, beyond the usual $2$-local models previously analyzed.

Finally, we investigate the role of the target Hamiltonian landscape geometry~\cite{BaldassiPNAS2016} in the effectiveness of digitized-QA vs QAOA for our model. This is done by artificially shuffling the classical energies associated to each spin configuration: despite the spectrum and the number of classical solutions being the same, a gap closure on the adiabatic path appears~\cite{BaldassiPNAS2018}, which has well-known detrimental effects on QA performance.
We show that also our QAOA implementation is affected by the gap closure, even though it still offers some advantages compared to digitized-QA.

The rest of the article is organized as follows:
in Sec.~\ref{sec:methods} we review the formulation of digitized-QA and QAOA, and introduce the model Hamiltonian for the binary perceptron. 
In Sec.~\ref{sec:results} we discuss our numerical results on the comparison of these two algorithms, along with the aforementioned qualitative features of the optimal solutions, such as smoothness and mild dependence on the training set. In Sec.~\ref{sec:landscape_geometry} we analyze the role of the problem landscape geometry, while in Sec.~\ref{sec:conclusions} we discuss possible extensions and generalizations of our work.

\section{Problem and methods} \label{sec:methods}

\subsection{Digitized-QA and QAOA}
In this section we summarize the main ingredients of a digitized-QA\cite{Nature_dQA,Mbeng_dQA_PRB2019} and QAOA, as a variational quantum algorithm~\cite{Mbeng_arXiv2019}.

In standard QA/AQC framework~\cite{Albash_RMP18} one constructs an interpolating Hamiltonian $\Ho(s) = s \Ho_{\target} + (1-s) \Ho_x$, where $\Ho_{\target}$ is the problem (or target) Hamiltonian ---
which depends on spin-$1/2$ Pauli operators and whose ground state we wish to find ---, while $\Ho_x = - \Gamma_0 \sum_j \PauliSigma^x_j$ is a transverse field term, allowing for quantum fluctuations.
For a classical minimization problem, as the case we are investigating, $\Ho_{\target}=\Ho_z$ is simply built from $\PauliSigma^z$ terms as in Eq.~\ref{eq:embedding}, thus it is diagonal in the computational basis~\cite{Nielsen_Chuang:book}, and its ground states encode the classical solutions; in contrast, for the case of quantum state preparation, $\Ho_{\target}$ contains further non-diagonal quantum fluctuation terms. 
An adiabatic dynamics is then pursued by slowly increasing $s(t)$ from $s(0)=0$ to $s(\tau)=1$ in a large total annealing time 
$\tau$, starting from the easily-prepared GS of $\Ho_x$:
\begin{equation}\label{eq:init_state}
    |+\rangle^{\otimes N} = \left( \frac{\upket+\downket}{\sqrt{2}}  \right)^{\otimes N}\;,
\end{equation}
where $\upket$ and $\downket$ denote the spin up/down eigenstates of $\PauliSigma^z$. 

The difficulty in the QA/AQC scheme is usually associated with the growing values of the annealing time $\tau$ required to adiabatically follow the instantaneous GS of $\Ho(s)$, possibly diverging in the thermodynamic limit, if the system crosses a critical point or, even worse, a first order phase transition~\cite{Zamponi_QA:review}.
While in principle the annealing schedule $s(t)$ can be chosen with some freedom, in practice optimizing the $s(t)$, by ``slowing down'' close to points where the spectral gap of $\Ho(s)$ closes, requires knowing such spectral information, a notoriously difficult problem~\cite{Cubitt_Nat2015}. Very often, a {\em linear} schedule $s(t)=t/\tau$ is assumed. 

The digitalization of the continuous-time QA/AQC dynamics is a very natural idea, conceptually relying on the Trotter split-up of non-commuting exponentials.
For the case of a classical combinatorial optimization problem, $\Ho_{\target}=\Ho_z$,  we would simply write:
\begin{equation}\label{eq:discrete_dynamics}
    \nep^{-i\frac{\Delta t}{\hbar} \Ho(s) } = \nep^{-i \beta \Ho_x} \nep^{-i \gamma \Ho_z} + \mathcal{O}\left( (\Delta t)^2 \right)  \;,
\end{equation}
with $\beta=(1-s) \Delta t/\hbar$ and $\gamma=s \Delta t/\hbar$ to lowest order in the Trotter-splitting.
Similarly, if $\Ho_{\target}$ is a combination of a $\PauliSigma^z$-part $\Ho_z$ and a $\PauliSigma^x$-part $\Ho_x$ --- as for the ground state preparation of an Ising model (or Ising spin glass) in a transverse field --- the same expression still holds, with suitable $\beta$ and $\gamma$ obtained by the Trotter split-up.

With the standard QA/AQC assumption of a linear annealing schedule $s(t)=t/\tau$, we would perform a digitized-QA by simply setting
$s_m=m/\Ptrot$ for $m=1\cdots \Ptrot$ where $\Ptrot$ is the number of {\em Trotter steps}, and $\Delta t_m=\Delta t$, hence a total annealing time $\tau=\Ptrot \Delta t$. 
This amounts to setting, in the case of a classical optimization problem, for all $m=1\cdots \Ptrot$:
\begin{equation} \label{eq:dQA_angles}
     \beta_m = (1-s_m) \frac{\Delta t}{\hbar} \;, \hspace{5mm}
     \gamma_m = s_m \frac{\Delta t}{\hbar} \;, \hspace{5mm} s_m=\frac{m}{\Ptrot} \;.
\end{equation}
The digitized-QA unitary evolution would hence be given by:
\begin{equation} \label{eq:QAOA_state}
|\psi_{\Ptrot}(\bbeta,\bgamma)\rangle =
\hat{U}(\beta_\Ptrot,\gamma_\Ptrot) \cdots \hat{U}(\beta_1,\gamma_1) |+\rangle^{\otimes N} \;,
\end{equation}
where $\bbeta=(\beta_1,\cdots\beta_{\Ptrot})$, $\bgamma=(\gamma_1,\cdots \gamma_{\Ptrot})$ and the $m$-th step evolution operator reads:
\begin{equation} \label{eq:U_m}
    \hat{U}(\beta_m,\gamma_m) = 
    \nep^{-i \beta_m \Ho_x} \nep^{-i \gamma_m \Ho_z} \;.
\end{equation}
This digitized-QA scheme allows for a single variational parameter, $\Delta t$ in Eq.~\eqref{eq:dQA_angles}, which we can optimize so as to minimize the variational energy:
\begin{equation} \label{eq:QAOA_en}
    E_\Ptrot \left(\bbeta, \bgamma\right) = \langle \psi_{\Ptrot}(\bbeta,\bgamma) | \Ho_{\target} | \psi_{\Ptrot}(\bbeta,\bgamma) \rangle \;,
\end{equation}
with $\Ho_{\target}=\Ho_z$ for a classical optimization.
Indeed, as discussed in Ref.~\cite{Mbeng_dQA_PRB2019}, a too-small value of $\Delta t$ produces small Trotter errors but also a short annealing time, while a too large $\Delta t$ is associated to large Trotter errors that make the final state rather inaccurate. Consequently, there is an optimal value of $\Delta t$ that one can choose in performing such a digitized-QA dynamics~\cite{Mbeng_dQA_PRB2019}.

The Quantum Approximate Optimization Algorithm (QAOA) by Farhi {\em et al.}~\cite{Farhi_arXiv2014} and more general
Variational Quantum Algorithms are based on the basic variational principle of Quantum Mechanics: a trial parameterized wavefunction is defined --- usually in terms of a quantum circuit --- with an objective function given by the target Hamiltonian average energy on the state.
One regards the quantum wavefunction parameters as {\em variational parameters} of the objective function in Eq.~\ref{eq:QAOA_en},
which we seek to minimize by some appropriate {\em classical} optimization scheme: hence the hybrid quantum-classical nature of such algorithms. 

In the QAOA case, to which we shall restrict our considerations from now on, the trial wavefunction has the same form as in Eqs.~\eqref{eq:QAOA_state}-\eqref{eq:U_m}, where now $\bbeta$ and $\bgamma$ are promoted to variational parameters for the quantum state, rather than fixed by a Trotter split-up as in digitized-QA.
The {\em pro} of such a variational scheme is that
the optimal energy at the global minimum $E_\Ptrot \left(\bbeta^*,\bgamma^* \right)$ is certainly a monotonically decreasing function of $\Ptrot$, which systematically improves
on any digitized-QA approach with the same $\Ptrot$.
The {\em cons} is that determining the global minimum $\left(\bbeta^*,\bgamma^* \right)$ is in general a non-trivial task,
since local optimization routines tend to get trapped into one of the many local minima of the $2\Ptrot$-dimensional search space, and the phenomenon of {\em barren plateaus}~\cite{McClean_NatCom2018}
can make the gradients of the objective function exponentially small in the number of spin variables. 

Let us finally stress that while the implementation of
$\nep^{-i \beta_{m} \Hx }$ requires a single layer of one-qubit gates, the gate decomposition (and thus the depth of the resulting quantum circuit) for the unitary $\nep^{-i \gamma_{m} \Hz} $ is strongly problem-dependent, and usually represents the true bottleneck for an actual implementation of this computational paradigm.
Let us mention, in this respect, there is a whole active field --- sometimes referred to as {\em quantum neural networks} --- where parameterized quantum circuits \cite{PQC_ML} with fixed gates but free variational parameters are studied, both from the point of view of their expressive power --- the ``expressivity'' of a parameterized state being the set states of the Hilbert space that it is able to represent --- as well as from the ease in finding good parameters, the so-called ``trainability''.
In these cases, the target Hamiltonian $\Ho_{\target}$ (or the $\Ho_z$ of the classical optimization problem) needs not be directly used in the construction the unitary gates.

\subsection{Perceptron model}
The perceptron represents the prototypical example of a single-layer binary classifier, first introduced decades ago by Rosenblatt~\cite{Rosenblatt_1957_6098}. It is still a subject of active research, both as the fundamental unit of classical artificial neural networks~\cite{Hertz:book} and as a potential candidate for simple realizations of quantum neural networks~\cite{Macchiavello, SCHULD2015660}.
\begin{figure}[!ht]
\centering
\begin{tikzpicture}[scale=1, every node/.style={scale=1}, square/.style={regular polygon, regular polygon sides=4}]
\draw (-4.6,2) node[right, text=black]{$\xi_1^{\mu}$}; 
\node[circle,draw=black,fill=black,inner sep=0pt,minimum size=4pt] at (-4,2) {};
\draw [line width = 0.5mm, draw=black, -] (-4,2) -- (0,0.2) node[right, black] {}; 
\node[circle,draw=black,fill=white,inner sep=0pt,minimum size=15pt] at (-3,0.2+0.75*1.8) {$\sigma_1$};
\node[circle,draw=black,fill=black,inner sep=0pt,minimum size=4pt] at (-4,1) {};
\draw (-4.6,1) node[right, text=black]{$\xi_2^{\mu}$}; 
\draw [line width = 0.5mm, draw=black, -] (-4,1) -- (0,0.1) node[right, black] {}; 
\node[circle,draw=black,fill=white,inner sep=0pt,minimum size=15pt,label=below:{}] (a) at (-3,0.1+0.75*0.9) {$\sigma_2$};
\node[circle,draw=black,fill=black,inner sep=0pt,minimum size=4pt] at (-4,-2) {};
\draw (-4.6,-2) node[right, text=black]{$\xi_N^{\mu}$}; 
\draw [line width = 0.5mm, draw=black, -] (-4,-2) -- (0,-0.1) node[right, black] {}; 
\node[circle,draw=black,fill=white,inner sep=0pt,minimum size=15pt,label=below:{}] (a) at (-3,-0.2-0.75*1.8) {$\sigma_{\scriptscriptstyle N}$};
\draw (-4.2,0.1) node[right, text=black]{$\vdots$}; 
\draw (-4.2,-0.8) node[right, text=black]{$\vdots$}; 
\draw [line width = 0.5mm, draw=black, -] (0,0) -- (2,0) node[right, black] {$\tau^{\mu}$};
\node[circle,draw=black,fill=blue,inner sep=0pt,minimum size=4pt] at (2,0) {};
\node[circle,draw=black,line width = 0.5mm,fill=white,inner sep=0pt,minimum size=40pt] at (0,0) {$\;\;\mathrm{sgn}(\bsigma\cdot\bxi^{\mu})\;\;$};
\end{tikzpicture}
\caption{Scheme of a Perceptron. Binary synaptic weights $\sigma_j$ have to be adjusted such that for given binary values $\xi_j^{\mu}$ in the $N$ input neurons, the scalar product $\bsigma\cdot\bxi^{\mu}=\sum_j \sigma_j \xi_j^{\mu}$ has a prescribed output sign $\tau^{\mu}$. 
Here $\mu=1\cdots M$, with $M=\alpha N$, labels the various input-output patterns.}
\label{fig:perceptron}
\end{figure}
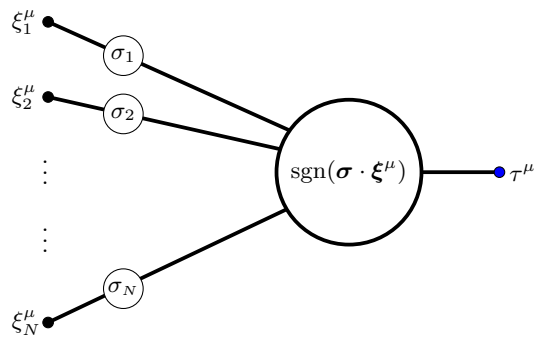

Following Ref.~\cite{BaldassiPNAS2018}, we address the problem of supervised learning of $M=\alpha N$ random patterns in a perceptron with $N$ neurons in the input layer: any configuration of the binary synaptic weights $\bsigma=\{\sigma_j\}\in \{-1,1\}^N$ correctly classifies a randomly generated pattern $\bxi^{\mu}=\{\xi_j^{\mu}\} \in\{-1,1\}^N$ into a prescribed binary label $\tau^{\mu}=\pm 1$ if $\mathrm{sgn}(\bsigma \cdot \bxi^{\mu})=\tau^{\mu}$,
see sketch in Fig.~\ref{fig:perceptron}.

During the learning phase, a given labeled data-set $\{\bxi^\mu,\tau^\mu\}_{\mu=1}^{M}$ is provided, and the task consists in finding the weight configurations $\bsigma$ such that all the patterns are correctly classified.
Since, by hypothesis, both the components of the patterns $\bxi^\mu$ and the labels $\tau^\mu$ are independent identically distributed (unbiased) Bernoulli random variables --- results of a fair coin flip --- we can assume without loss of generality the labels $\tau^{\mu}$ to be all equal $+1$.

The search problem is immediately reformulated as a minimization problem of a suitable cost function.
The underlying idea is to associate a positive energy cost for every pattern incorrectly classified.
The exact solutions to the classification problem are the zero-energy configurations $\bsigma^*$ of the cost function. 
Let us define 
\begin{equation} \label{eq:m_mu}
m_{\mu} = \frac{1}{\sqrt{N}} \sum_{j=1}^N \sigma_j \xi^{\mu}_j    
\end{equation}
to be the overlap between the spin configuration $\bsigma$ and the $\mu$-th pattern $\bxi^{\mu}$, normalized in such a way that, upon averaging over the random patterns, one gets $\overline{m_{\mu}^2} = 1$.
A possible definition of the cost function is:
\begin{equation} \label{eq:cost_function}
E_{\ncost} \left(\{ \sigma_j \}\right) := \sum_{\mu=1}^{M} |m_{\mu}|^{\ncost} \, \Theta\left(-m_{\mu}\right) \;,
\end{equation}
where $\Theta(x)=(1+\mathrm{sgn}(x))/2$ is the Heaviside step function. The energy cost for an incorrect classification of a pattern, $m_{\mu}<0$, is simply $+1$ if $\ncost=0$, or proportional to the error, $|m_{\mu}|$, if $\ncost=1$. 
We remark that for both values of $\ncost=0, 1$ the cost function yields the same global minima $\bsigma^*$ at zero-energy (exact solutions), while the energy landscapes and the local minima are generally different. 

Finding optimal solutions $\bsigma^*$ where $E_{\ncost}(\bsigma^*)=0$ is a hard optimization problem for either choice of $\ncost=0,1$: it has been shown that the energy landscape, in the limit of large $N$ and for $M=\alpha N$ with  $\alpha<\alpha_c\approx0.83$~\cite{Mezard_perceptron}, is characterized by an exponential number of zero-energy solutions and local minima. 
The latter play the role of metastable states for classical stochastic search algorithms, such as Simulated Annealing (SA)~\cite{Horner}, which typically get stuck, for large $N$, in one of these local minima, with extensive energy costs (of the order $\mathcal{O}(N)$).

More recently, further insight has been gained about the geometrical structure of the ground states~\cite{BaldassiPNAS2016}. Very schematically, exponentially rare regions where the ground states are dense exist: these regions are composed of an \emph{exponentially large} number of zero-energy solutions at extensive but relatively small Hamming distances, thus possessing very high local entropy. 
Despite being exponentially rare, these dense regions might be particularly well-suited for making predictions after the supervision is carried out, since they are less likely to fit noise (small generalization error).

It has been conjectured and shown in Ref.~\cite{BaldassiPNAS2018} --- with analytical and numerical evidence --- that quantum fluctuations, as encoded by a Path-Integral Monte Carlo simulated QA~\cite{Santoro_SCI02}, are particularly effective in exploring these ``dense'' regions. 
One of the aims of this work is to provide numerical evidence of enhanced effectiveness of QAOA over QA for small-size perceptron instances, where unitary evolutions are computationally feasible to directly compare QA and QAOA.  

The standard quantum mapping of the binary synaptic weights consists in promoting the classical spins $\sigma_j$ to quantum spin-$1/2$ Pauli operators $\PauliSigma^z_j$, as a special case of the procedure schematized in Eq.~\ref{eq:embedding}.
Let us remark that this standard mapping is not the only possibility to encode classical bits into a quantum setup.
A recent work~\cite{Macchiavello} has implemented a quantum version of the perceptron model within the so-called {\em amplitude encoding}:
such a scheme is in principle very efficient in terms of memory resources, as it requires $\log_2 N$ quantum spins to represent $N$ classical spins, but it pays the price of an exponentially large number of quantum gates necessary for the state preparation~\cite{Macchiavello}.
In our study, we focus on leveraging quantum fluctuations to train a \emph{classical} perceptron, rather than implementing a quantum version of it: as a natural choice, in the following, we proceed with the standard encoding $\sigma_j\to \PauliSigma^z_j$.

The target Hamiltonian associated to the perceptron is then given by
\begin{equation} \label{eq:Ham_perceptron}
\Ho_{\target} = 
E_{\ncost}\left(\{\PauliSigma^z_j\}\right)  \;.
\end{equation}
The operator $\Ho_z=\Ho_{\target}$ is also used in the QAOA variational state in Eqs.~\eqref{eq:QAOA_state}-\eqref{eq:U_m}.
We remark that this classical Hamiltonian is highly non-local due to the presence of the Heaviside $\Theta$ in the activation function --- in principle implying all possible interactions among spins, up to $N$-body terms, 
and hence hardly implementable on an actual device. 
%
As anticipated above, the QAOA-like {\em Ansatz}  allows for some flexibility in the choice of the $\Ho_z$ term appearing in the variational state in Eqs.~\eqref{eq:QAOA_state}-\eqref{eq:U_m}, with the possibility of replacing it by a simpler set of quantum gates, a point that we will further discuss in Sec.~\ref{sec:conclusions}.


\section{Results on the Perceptron model} \label{sec:results}

In order to perform a fair comparison of QAOA against QA (in its digitized form), we consider a set of $10$ instances of the perceptron problem for $N=21$ spins, which were previously analyzed in 
\cite{BaldassiPNAS2018}(SI).
%
For each instance, we aim at classifying correctly a training set of $M=17$ patterns, corresponding to $\alpha=\frac{M}{N}\approx0.81$, close to the critical value $\alpha_c\approx 0.83$, valid in the thermodynamic limit $N\to\infty$, beyond which zero-energy solutions may no longer exist.
%
%

Following Ref.~\cite{BaldassiPNAS2018}(SI), these instances were obtained by randomly generating $450$ candidate \emph{training sets samples}, and: i) keeping only those with a sufficiently large number  of solutions ($>21$, thus hinting at a non-convex optimization problem);  
%
ii) keeping only the instances for which SA failed to reach good approximate solutions. 
From here on, we shall refer to a perceptron instance characterized by a specific randomly-generated training set simply as \emph{sample}.


\subsection{Optimal digitized-QA}

\begin{figure}[ht]
\centering
\includegraphics[width=0.485\textwidth]{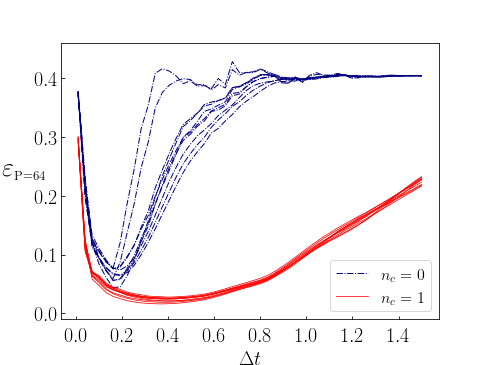}
\caption{
The one-dimensional landscape, versus $\Delta t$, of the variational energy density $\varepsilon_{\Ptrot}(\bbeta,\bgamma)$, Eq.~\eqref{eq:epsilon_P}, for the digitized-QA parameters, Eq.~\eqref{eq:dQA_angles}.  
All samples examined and both choices of $\ncost=0,1$ are shown. The qualitative features of the landscape, and in particular the position of global minima, show mild sample-to-sample variability.   
}
\label{fig:dQA_summary}
\end{figure}

The digitized-QA (dQA) parameters $\bbeta$ and $\bgamma$ in Eq.~\eqref{eq:dQA_angles} contain $\Delta t$ as a single variational parameter, which we can use to minimize the variational energy density:
\begin{equation} \label{eq:epsilon_P}
    \varepsilon_{\Ptrot}(\bbeta, \bgamma)  
    = \frac{1}{N} E_\Ptrot (\bbeta, \bgamma) \;.
\end{equation}
%
Empirically, for any value of $\Ptrot$, a single global minimum for $\Delta t$ emerges very clearly for each sample. 
Fig.~\ref{fig:dQA_summary} shows $\varepsilon_{\Ptrot}$ for $\Ptrot=64$ and the two different cost-function choices, $\ncost=0$ and $\ncost=1$, versus $\Delta t$.

As previously discussed, see also \cite{Mbeng_dQA_PRB2019}, the rationale behind the presence of an optimal $\Delta t$ is simple:  Essentially, by increasing $\Delta t$ at fixed $\Ptrot$, $\varepsilon_{\Ptrot}$ initially decreases, because we are allowing for a longer annealing time $\tau=\Ptrot\Delta t$; upon further increase of $\Delta t$, however, Trotter errors start to spoil the results, and lead, eventually, to a noise-dominated regime.
Remarkably, the $\Delta t$-landscape and optimal value 
depend significantly only on the choice of $\ncost=0,1$ for the cost function in Eq.~\eqref{eq:cost_function}, while much smaller sample-to-sample variations are present.

This is a first hint of general qualitative features of the model, independent of the specific sample under consideration. 
Additional numerical evidence on digitized-QA is reported in Appendix~\ref{app:randomized_dQA}.

\subsection{Smooth QAOA solutions} 
%

\begin{figure*}[ht]
\centering
\includegraphics[height=0.28\textwidth]{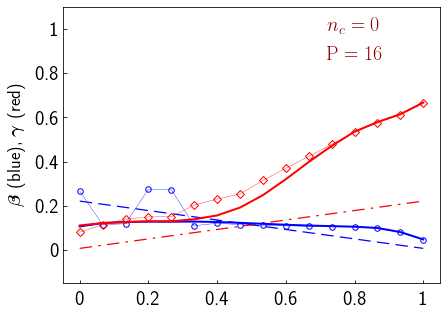}
\includegraphics[height=0.28\textwidth]{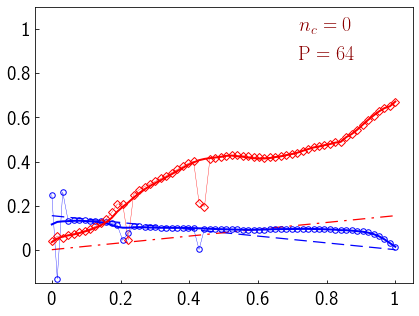}
\\
\includegraphics[height=0.3027\textwidth]{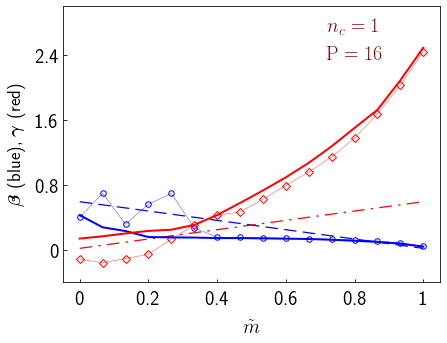}
\includegraphics[height=0.3027\textwidth]{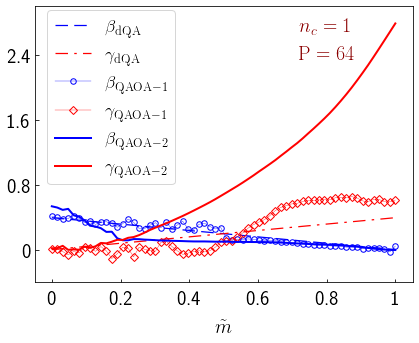} 
\caption{
Results for the optimal digitized-QA protocol (dashed and dash-dotted straight lines), QAOA-1 (open symbols with dotted lines), and QAOA-2 (solid lines), for $\Ptrot=16, 64$ (left to right) and for $\ncost=0, 1$ (top to bottom). We adopt a uniform $x$-axis scale in terms of $\tilde{m} = (m-1)/(\Ptrot-1) \in [0,1]$.
}
\label{fig:protocols_sample_one}
\end{figure*}
%
In practical implementations of QAOA, both the choice of the classical optimization algorithm --- local (gradient-based), versus global --- and of the starting point for the optimization routine can be relevant, particularly when the dimensionality $2 \Ptrot$ of the search space grows, making the optimization harder. 
%
The simplest approach would be to use a local gradient-based algorithm starting from a random initialization of the parameters, but in practice this is often ineffective, due to the presence of many local low-quality minima. Moreover, recent evidence proved the ubiquitous existence of barren plateaus for Variational Quantum Algorithms~\cite{McClean_NatCom2018, BP_shallow} --- i.e., gradients of the objective function that are exponentially vanishing in the system size $N$, when the landscape is sampled at random points --- a phenomenon that calls for effective warm-start strategies, further discouraging random initialization: we shall not adopt it here. 

Effective heuristic warm-start strategies have been proposed, which are based on iterative procedures in $\Ptrot$ and empirically yield far-better quality results than a random start.
These strategies are based on the fact that --- heuristically, and for the problems addressed so far --- some optimal or quasi-optimal solutions $(\bbeta,\bgamma)$ appear to be {\em smooth} when considered versus the Trotter number $m$, albeit sometimes strongly departing from the linear-choice of Eq.~\eqref{eq:dQA_angles}. Smoothness, however, does not come automatically and for free: you have in some way to look for it, either by adopting Fourier-based algorithms~\cite{Zhou_PRX2020}, or by interpolating from previous smaller $\Ptrot$ solutions ~\cite{Zhou_PRX2020,Mbeng_arXiv2019}.  

In this work we adopt the following strategy.
Let us denote by $(\bbeta^{\dQA},\bgamma^{\dQA})$ the optimal linear-choice that a digitized-QA provides, as discussed above. 
Using $(\bbeta^{\dQA},\bgamma^{\dQA})$ as a starting point for a Broyden-Fletcher-Goldfard-Shanno (BFGS) optimization algorithm~\cite{Nocedal_book2006, 2020SciPy}, we find a minimum, denoted by $(\bbeta^{(1)},\bgamma^{(1)})$, which is often 
``close'' to be a smooth curve, with occasional high-frequency localised oscillations of the optimal parameters.
To enforce smoothness, we apply a smoothing procedure to 
$(\bbeta^{(1)},\bgamma^{(1)})$ and restart a second BFGS optimization, leading to a final solution 
$(\bbeta^{(2)},\bgamma^{(2)})$ which is found to be {\em smooth}, and providing a systematically better variational mininum as compared to $(\bbeta^{(1)},\bgamma^{(1)})$. Schematically, here is the procedure adopted: 
\begin{widetext}
\begin{equation} \label{eq:scheme_QAOA}
(\bbeta^{\dQA},\bgamma^{\dQA}) \to \framebox[1.1 \width]{BFGS optim.} \stackrel{\scriptscriptstyle \mathrm{QAOA-1}}{\longrightarrow} (\bbeta^{(1)},\bgamma^{(1)}) 
\to \framebox[1.1 \width]{Smoothing + BFGS optim.} \stackrel{\scriptscriptstyle \mathrm{QAOA-2}}{\longrightarrow} (\bbeta^{(2)},\bgamma^{(2)})\;.
\end{equation}
\end{widetext}
Alternatively, a smooth $(\bbeta^{(2)},\bgamma^{(2)})$ solution could be obtained by interpolation~\cite{Zhou_PRX2020, Mbeng_arXiv2019} of a previous lower-$\Ptrot$ smooth solution. 

We now move to illustrate our results in more detail.
We performed digitized-QA and QAOA classical simulations for both $\ncost=0,1$: in this framework, we computed the exact QAOA final state in Eq.~\ref{eq:QAOA_state} and the corresponding variational energy in Eq.~\ref{eq:QAOA_en} by applying the algebra of Quantum Mechanics.
While this analysis was carried out for all samples under study, we now focus, for the sake of clarity, on a single sample (or training set), with similar comments and results applying to all samples.  

Fig.~\ref{fig:protocols_sample_one} illustrates the results obtained for this sample. We show two representative values of $\Ptrot=16,64$ (left to right) and both energy-cost functions $\ncost=0,1$ (top to bottom). 
The dashed straight lines denote the optimal $(\bbeta^{\dQA},\bgamma^{\dQA})$ solutions.
The empty symbols denote the optimal $(\bbeta^{(1)},\bgamma^{(1)})$ solution obtained by a BFGS minimization starting from $(\bbeta^{\dQA},\bgamma^{\dQA})$: our ``first shot of QAOA'', labelled as QAOA-1.
Notice the irregularities over an overall ``smooth'' behavior, particularly evident for $\Ptrot=64$ both at $\ncost=0$, where irregularities are quite localised, and for $\ncost=1$. 

For $\ncost=0$ we apply a smoothing procedure, 
and start a ``second shot'' of QAOA simply as summarized in Eq.~\ref{eq:scheme_QAOA}. 
For $\ncost=1$ the irregularities of the QAOA-1 solution are more diffuse, and the procedure was slightly modified: we run the second BFGS local minimization from a warm-start point, obtained by interpolation from a smoothed $\Ptrot=16$ solution, in power-of-two steps, hence from $\Ptrot=16\to 32 \to 64 \cdots$. 
In both cases, the resulting smooth solutions $(\bbeta^{(2)},\bgamma^{(2)})$ are labelled as QAOA-2 and denoted by solid lines.  
In Appendix~\ref{app:details} we summarize few more technical details concerning these two procedures to single out QAOA-2 smooth solutions; however --- as discussed in the next section --- we anticipate that these are not particularly crucial: once a smooth solution for a \emph{single} training set sample is found, there is no need to repeat the whole procedure for other samples.

\begin{figure}[ht]
\centering
\includegraphics[width=0.475\textwidth]{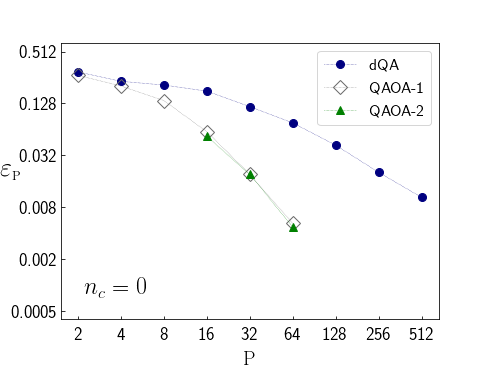}
\includegraphics[width=0.475\textwidth]{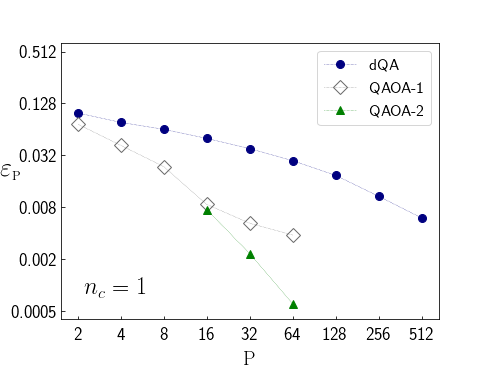}
\caption{Comparison of variational energy density minima for digitized-QA, QAOA-$1$ and QAOA-$2$ for a specific sample of random patterns, with $\ncost=0$ (top) and $\ncost=1$ (bottom). Notice that both axes have logarithmic scales: QAOA-$1$ outperforms digitized-QA, especially for large values of $\Ptrot$. This gain can be further enhanced with the smooth QAOA-$2$ solution (see main text).}
\label{fig:dQA_vs_QAOA}
\end{figure}

In Fig.~\ref{fig:dQA_vs_QAOA} we compare the obtained minima of $\varepsilon_{\Ptrot}$, Eq.~\eqref{eq:epsilon_P}, for the optimal digitized-QA, QAOA-$1$ and QAOA-$2$ protocols.
The results show a striking gain by applying QAOA (notice the logarithmic scales on both axes), both for $\ncost=0$ and $\ncost=1$.
Moreover, as anticipated, the smooth QAOA-$2$ protocol yields better results than the starting QAOA-$1$ protocols. 
%
These results hold true for all randomly generated training set samples. 
In light of these findings, the QAOA-1 solutions 
can be interpreted as local minima --- where the classical BFGS optimization gets trapped --- systematically of lower quality than the corresponding smoothed QAOA-$2$ protocols. 

%

\subsection{Transferability of a smooth Ansatz}

\begin{figure*}[ht]
\centering
\includegraphics[height=0.28\textwidth]{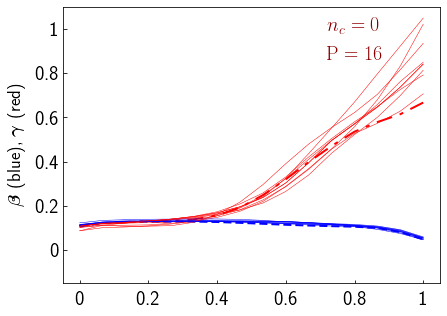}
\includegraphics[height=0.28\textwidth]{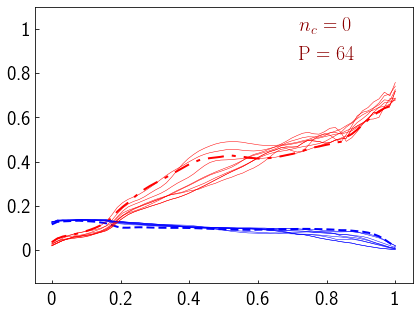} \\
\includegraphics[height=0.3027\textwidth]{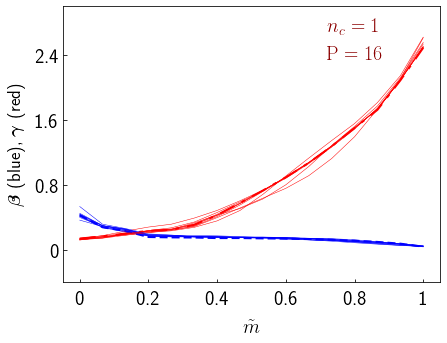}
\includegraphics[height=0.3027\textwidth]{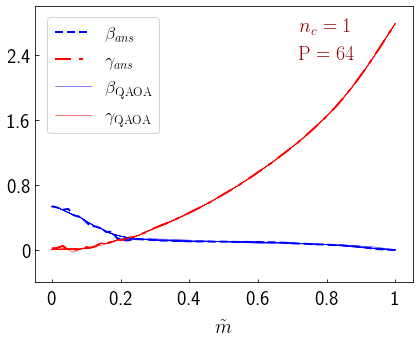}
\caption{
QAOA optimal protocols for all tested samples, as obtained by a BFGS minimization starting from the same {\em smooth Ansatz} $(\bbeta^{\ans},\bgamma^{\ans})$, corresponding to smooth optimal schedules for the first sample (see main text). Data shown for $\Ptrot=16, 64$ (left to right) and for $\ncost=0,1$ (top to bottom). 
We adopt a uniform $x$-axis scale in terms of $\tilde{m} = (m-1)/(\Ptrot-1)\in [0,1]$. The qualitative similarity of QAOA solutions for different samples is remarkable, particularly for $\ncost=1$, where optimal protocols for different samples are almost indistinguishable.
}
\label{fig:smooth_protocols}
\end{figure*}

The procedure described above to obtain smooth solutions is elaborate and time-consuming. Luckily enough, it does not need to be repeated for each different training set sample.

Indeed, a {\em leitmotif} of recent literature on QAOA applications are \emph{concentration} effects of the energy landscape: for any given $\Ptrot$, typical instances of the same problem often yield similar QAOA energy landscapes. 
When this is the case, a QAOA solution for the first instance often serves as an excellent warm-start for a local optimization of other instances, with a significant reduction of computational costs.

While this result can be formally understood, sometimes, in terms of light-cone correlations spreading, {\em e.g.} for MaxCut problems on regular graphs for $\Ptrot\ll N$~\cite{brandao2018, galda2021transferability}, a growing body of numerical evidence hints at concentration effects in different regimes (large $\Ptrot$) and for different models, often without a complete formal understanding.
A remarkable exception is offered by~\cite{farhi2020quantum}, where the authors prove concentration for the Sherrington-Kirkpatrick (SK) model in the large-$N$ limit, although the infinite range of 2-body interactions hinders an intuitive understanding in terms of light-cone spreading of correlations.

In the following, we show numerically that similar concepts apply to the perceptron model, which does not even admit a $k$-local cost function: QAOA smooth solutions are transferable for different instances of the perceptron model (i.e. different training set samples).
%
%
%
%
To show this, we adopt the following strategy: 
for any fixed value of $\Ptrot$ and for $\ncost=0,1$, separately, we consider QAOA-$2$ optimal angles $(\bbeta^{(2)},\bgamma^{(2)})$ for our first sample and take them as a {\em smooth model-dependent Ansatz} $(\bbeta^{\ans},\bgamma^{\ans})$ 
used as initial point for a BFGS-minimization of a {\em different} sample (training set).
We find in this way, as illustrated in Fig~\ref{fig:smooth_protocols}, {\em smooth} optimal solutions for all other samples.
Remarkably, these smooth solutions are qualitatively coincident with the $(\bbeta^{(2)},\bgamma^{(2)})$ solutions one would construct for that sample by using the QAOA-2 procedure outlined previously.
%
%
From a practical standpoint, by starting from the smooth {\em Ansatz} $(\bbeta^{\ans},\bgamma^{\ans})$, the convergence of the BFGS-optimization is always much faster. 

We remark, finally, that a certain variability of the curves in Fig.~\ref{fig:smooth_protocols} is likely to be ascribed to the small value of $N=21$ we used, while a genuine concentration might emerge for large-$N$, as in the SK case~\cite{farhi2020quantum}.

\section{The role of the problem landscape geometry}
\label{sec:landscape_geometry}
\begin{figure*}[ht]
\centering
\includegraphics[width=0.485\textwidth]{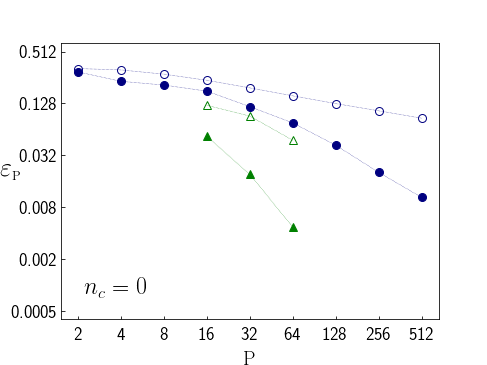}
\includegraphics[width=0.485\textwidth]{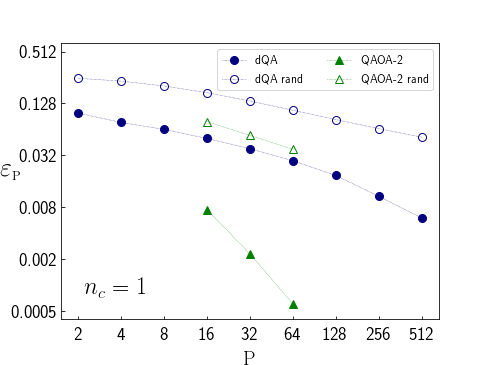}
\caption{Variational energy density minima for digitized-QA and QAOA-2, comparing results for the original sample (the same as in Fig.~\ref{fig:dQA_vs_QAOA}, full symbols) and that obtained by a {\em randomization} of its energy spectrum (empty symbols), for both $\ncost=0,1$.
}
\label{fig:dQA_vs_QAOA_randomized}
\end{figure*}
As discussed in Ref.~\cite{BaldassiPNAS2018}, quantum fluctuations are particularly efficient in exploring exponentially rare \emph{dense} regions of the classical configurations landscape, characterized by a large number of classical solutions clustering within relatively small Hamming distance. 
%
%
The geometrical structure of the landscape is linked to the good performance of QA: the instantaneous spectral gap that the quantum annealing dynamics ``sees'' only closes when approaching the end of the protocol  $s\to 1$, where $\Ho(s)\to \Ho_{\target}$, due to the degeneracy of the classical solutions.

We can scramble this geometrical structure by shuffling the classical energies associated to each configuration, while keeping the spectrum unchanged. Ref.~\cite{BaldassiPNAS2018} showed that this scrambling is very detrimental to QA: it causes a sharp drop of the instantaneous spectral gap at a finite $s_c<1$, the usual bottleneck of QA, and a drastic worsening of the QA performance. 

It is natural to ask to what extent a QAOA protocol is able to cope with such a scrambling of the landscape geometry and with the associated ultra narrow spectral gap (avoided level-crossing): after all, QAOA is based on the variational principle, rather than on the adiabatic theorem.
However, smooth QAOA solutions might be a signal of an ``optimal adiabatic schedule''~\cite{Mbeng_arXiv2019}, and this might suggest a worse performance.

To answer such a question, we have adopted the following strategy. 
For each sample considered, we generate a corresponding {\em randomized sample}, by reshuffling the classical energies associated to each configuration $\bsigma$, so as to retain the same classical energy spectrum, while completely destroying any geometrical feature of the energy landscape.
%
We then proceed along the lines of Sec.~\ref{sec:results}: by starting from an optimal-$\Delta t$ digitized-QA solution for the randomized samples, we run QAOA and compare the minima of the variational energy density for the two methods.
Remarkably, it is again possible to single out {\em smooth}  QAOA-2 solutions, by following the scheme outlined in Eq.~\ref{eq:scheme_QAOA} (see details in Appendix~\ref{app:details}), and these smooth solutions are {\em transferable} among different randomized samples. 
%
In Appendix~\ref{app:randomized_dQA}, we carry out a closer comparison between digitized-QA optimal-$\Delta t$ values for the two cases of original vs randomized samples.

Fig.~\ref{fig:dQA_vs_QAOA_randomized} compares the minimized variational energy density obtained from digitized-QA and QAOA-2 for the randomized version of the sample reported in Fig.~\ref{fig:dQA_vs_QAOA};  the original digitized-QA and QAOA-2 results are also reported, for a direct comparison. 
Two main remarks are worthwhile: 1) the QAOA-2 solutions for the randomized sample considerably improve on the corresponding digitized-QA results, especially for $\ncost=0$, where they become comparable to the original sample digitized-QA results; 2) the quality of these QAOA-2 solutions is much worse than the original sample QAOA-2 solution, witnessing a degradation of the performance. 
These comments apply to all the samples examined. 

\begin{figure*}[ht]
\centering
\includegraphics[height=0.35\textwidth]{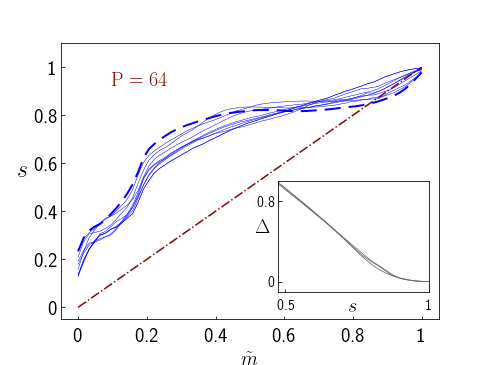}
\includegraphics[height=0.35\textwidth]{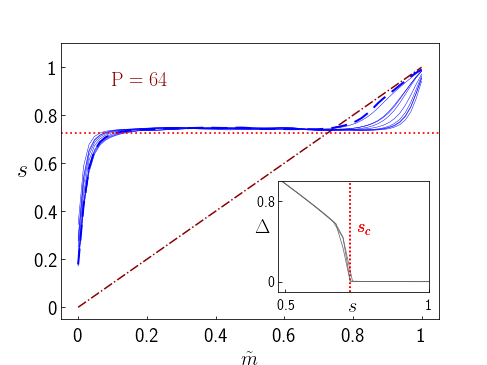}
\caption{Plot of the QAOA-2 smooth protocols for $\Ptrot=64$, in terms of $s_m=\gamma_m/(\gamma_m+\beta_m)$ for both the original (left) and the randomized samples (right), in the case $\ncost=0$. In both figures, the thick dashed line represents the transferable smooth \emph{Ansatz} obtained by a detailed study of the first sample, while data for other samples is obtained by exploiting this transferability result. As a visual reference, we also plot the digitized-QA linear interpolation from $s=0$ to $s=1$.
In the insets we show the instantaneous spectral gap $\Delta(s)=E_\text{ex}(s)-E_\text{gs}(s)$: the gap starts at $\Delta(s=0)=2$ (single spin-flip excitation of $\Hx$) and it vanishes for $s=1$, due to the degeneracy of $\Hz$, while negligible sample-to-sample variability is observed.
Remarkably, $\Delta(s)$ shows a sharp drop around $s_c=0.725$ for the randomized samples, and a wide plateau is highlighted in the corresponding smooth optimal schedules by a red dotted horizontal line.
Similar results and comments apply for $\ncost=1$.
}
\label{fig:gap_protocols_randomized}
\end{figure*}

To better understand the basic mechanism behind such a degradation of performance, we plot in Fig.~\ref{fig:gap_protocols_randomized} the QAOA-2 smooth protocols for $\Ptrot=64$,
re-expressed in terms of $s_m=\gamma_m/(\gamma_m+\beta_m)$, a parameter which in digitized-QA linearly interpolates from $s=0$ to $s=1$ during the annealing process, see Eq.~\eqref{eq:dQA_angles}. 
The two figures correspond to the original (left) and the randomized samples (right), in the case $\ncost=0$.
These optimal schedules should be compared with the instantaneous spectral gap $\Delta(s)=E_\text{ex}(s)-E_\text{gs}(s)$, which is plotted in the inset for both cases. 
We observe that the instantaneous gap of randomized samples displays, as expected~\cite{BaldassiPNAS2018}(SI), an avoided level-crossing close to the numerical value $s_c=0.725$. Correspondingly, close to this value, the optimal schedule parameter $s_m$ shows a wide marked plateau, particularly evident for large values of $\Ptrot$: this is reminiscent of a ``slowing down'' of the annealing near the gap closure, a kind of ``optimal adiabatic schedule''~\cite{Mbeng_arXiv2019}, unfortunately unable to fully overcome the basic limitations of the adiabatic mechanism. 

As a final remark, we mention that nothing guarantees that our QAOA-2 smooth solutions are the true global minimum in the $2\Ptrot$-dimensional energy landscape: in principle, there might be other better performing QAOA-protocols, similar to the ``short-cut to adiabaticity'' schedules found in some hard-instances of 3-MaxCut~\cite{Zhou_PRX2020}.

\section{Discussion and conclusions}
\label{sec:conclusions}
In our work we provided first encouraging evidence for the potential applicability and effectiveness of digitized-QA and QAOA methods in the realm of hard classical optimization problems with highly non-local Hamiltonians, well-beyond the usual $2$-local models considered both in MaxCut problems and quantum spin chains. 

Moreover, we devised an optimization scheme that leverages the transferability of optimal \emph{smooth} QAOA schedules among different instances of the same problem. 
These findings are qualitatively similar to previous results on landscape concentration effects and might be further investigated, possibly leading to analytical results in the large $N$ limit, as in Ref.~\cite{farhi2020quantum}.

Inspired by the results of Ref.~\cite{BaldassiPNAS2018}, we also investigated the role of the landscape geometry and associated small spectral gaps for QAOA in our model: while still providing some advantage vs an optimized-$\Delta t$ linear schedule QA, QAOA seems to perform a kind of ``optimal adiabatic schedule'', unable to fully overcome the basic limitations of the adiabatic mechanism.

Concerning future developments, it is an open research line to tailor smart approximation schemes to implement efficiently such complex Hamiltonians on actual quantum hardware --- paving the way to experiments beyond the classical simulation capabilities.
Nevertheless, we highlight that similar non-trivial tasks need to be solved also for the implementation of such models on a standard Quantum Annealer~\cite{BaldassiPNAS2018}.

In this respect, one possibility is a further extension of QAOA, proposed in Ref.~\cite{QAlternatingOA}. 
The main idea is that, by looking at Eqs.~\eqref{eq:QAOA_state}-\eqref{eq:U_m}, one can redefine the $2\Ptrot$-dimensional variational energy as follows: 1) keep the same exact classical $\Ho_{\target}$ in the objective function expectation value and 2) redefine the diagonal unitary (which is quantum computationally hard) generated by $\Hz$ --- {\em e.g.}, by using a simplified version of it, encoding some minimal information on the problem. As an additional possibility, one can even redefine the {\em mixing} unitary generated by $\Hx$, replacing it with another operator inducing tailored quantum fluctuations.
One would then simply repeat the QAOA prescription, but with a different --- and possibly much more efficient --- quantum gate implementation. In contrast, note that the expectation value, in a realistic quantum device, is simply estimated by computing the sample mean of $\Ho_{\target}=\Ho_z$ over a set of strings --- obtained by repeatedly preparing the QAOA state and measuring it on the computational basis: each evaluation of $\Ho_{\target}$ for a classical configuration $\bsigma$ is a trivial task, regardless of its non-locality. 

A simple attempt in this direction, which we have tried, is the following. Rather than using $\Ho_z=\Ho_{\target}$ in the quantum gates $\nep^{-i\gamma_m \Ho_z}$, we use the Sherrington-Kirkpatrick (SK) model Hamiltonian, which derives from taking the quadratic approximation $(|m_{\mu}|) \, \Theta(-m_{\mu})\to - m_{\mu} + m_{\mu}^2$ in the $\ncost=1$ cost function in Eq.~\eqref{eq:cost_function}, which, upon using Eq.~\eqref{eq:m_mu}, leads to:
%
\begin{equation}
\Ho_z 
= - \sum_{j=1}^N h_j \PauliSigma^z_j + \sum_{j\neq j'} J_{jj'} \PauliSigma^z_j \PauliSigma^z_{j'} \;,
\end{equation}
where $h_j = \frac{1}{\sqrt{N}} \sum_{\mu=1}^M \xi^{\mu}_j$  
is a local field provided by all the input patterns at site $j$, while
$J_{jj'} = \frac{1}{N} \sum_{\mu=1}^M \xi^{\mu}_j \xi^{\mu}_{j'}$
is the standard Hebbian-rule coupling~\cite{Hertz:book}.
Unfortunately, such a choice of $\Ho_z$ appears to dramatically decrease the performance of the QAOA algorithm.
A possibly smarter choice to improve performance, while maintaining a Hamiltonian with only two-body interactions, could be to search for an optimal SK model, either by iteratively identifying the optimal $J$ and $h$ parameters or by defining an appropriate inverse Ising model \cite{nguyen2017inverse}. These options will be the subject of future research.
%

An efficient implementation of an effective parameterized quantum circuit might pose new interesting questions, concerning not only the presence of QAOA smooth protocols and their transferability, but also their robustness to shot-noise (due to finite-sample mean estimates of the variational energy) or gate errors, effects worth studying only once an efficient gate implementation is found.

Finally, the role of classical energy landscape geometry --- as in ANNs~\cite{BaldassiPNAS2016} --- might be crucial for the effectiveness of QAOA or similar computational schemes to perform optimization tasks.
In this framework, a promising route might be to extend this work to the supervision of more complex architectures of ANNs, as well as consider training sets with correlations (data with a structure).

Interestingly, a recent work~\cite{Carrasquilla_NatMI_2021} has shown that purely classical strategies using recurrent neural networks (RNN) --- denoted as `` variational neural annealing '' --- can improve significantly on Simulated Annealing and simulated-QA on standard benchmark problems. 
Our study confirms that the perceptron model appears to belong to a class of problems where quantum effects 
efficiently exploit the local geometry of the landscape, much better than Simulated Annealing: 
it is therefore an ideal candidate to test and benchmark the variational neural annealing strategies of Ref.~\cite{Carrasquilla_NatMI_2021} against competing quantum algorithms.


\section*{Acknowledgments} 
We thank M. Collura and G. Lami for stimulating discussions. The research was partly supported by EU Horizon 2020 under ERC-ULTRADISS, Grant Agreement No. 834402.
GES acknowledges that his research has been conducted within the framework of the Trieste Institute for Theoretical Quantum Technologies (TQT).

\appendix

\section{QAOA-2 practical implementation} \label{app:details}

As mentioned in the main text, two slightly different procedures are adopted in order to single out smooth optimal QAOA-2 solutions.

In fact, some qualitative differences arise between QAOA-1 results for $\ncost=0$ and $\ncost=1$, which are visible in Fig.~\ref{fig:protocols_sample_one} (empty symbols) for the first training set sample, but are present for all the other samples in exam. Concisely, we observe that for $\ncost=0$ the QAOA-1 optimal parameters $(\bbeta^{(1)},\bgamma^{(1)})$ are noticeably different from $(\bbeta^{\dQA},\bgamma^{\dQA})$ for all values of $P$, and the high-frequency oscillations are either completely absent or well-localized on top of the smooth solutions.
This observation motivates the original QAOA-2 procedure outlined in Eq.~\ref{eq:scheme_QAOA} (smoothing, second BFGS minimization), which is applied straightforwardly e.g. for $P=16, 64$, yielding the smooth $(\bbeta^{(2)},\bgamma^{(2)})$ protocols in Fig.~\ref{fig:protocols_sample_one} (solid curves).

On the contrary, for $\ncost=1$, we observe the same qualitative features e.g. for $P=16$, whereas for larger values such as $P=32, 64$ the QAOA-1 solutions seem to get ``trapped'' in a neighborhood of $(\bbeta^{\dQA},\bgamma^{\dQA})$, also displaying more extended high-frequency oscillations in the optimal parameters $(\bbeta^{(1)},\bgamma^{(1)})$.
This numerical evidence calls for a slightly different approach: we simply apply Eq.~\ref{eq:scheme_QAOA} prescription only for $\Ptrot=16$, and we find smooth solutions for larger values $\Ptrot'$ using an iterative procedure: for each $\Ptrot'>16$, we determine the new starting point for BFGS minimization by interpolating on the optimal smooth curve found for the previous value of $\Ptrot$. We implement this procedure in power-of-two steps, hence from $\Ptrot=16\to 32 \to 64$, but we expect to obtain similar results e.g. by means of a linear increment in $P$ at each iteration. 

Consistently with our intuition, one can check in Fig.~\ref{fig:dQA_vs_QAOA} that QAOA-2 offers a noticeable improvement for $\ncost=1$, since the solutions for $P=32,64$ have now ``escaped'' the digitized-QA qualitative shape.

We remark that these details --- concerning only the technical implementation of our QAOA-2 framework --- do not affect our central message, as reported in the main text: for each sample in exam, QAOA hints at the presence of a smooth solution that systematically outperforms (optimal-$\Delta t$) digitized-QA, as shown in Fig.~\ref{fig:dQA_vs_QAOA}. This QAOA-1 solution is sometimes affected by the presence of high-frequency oscillations which can be smoothed out without spoiling the result: on the contrary, QAOA-2 is systematically (albeit sometimes negligibly) outperforming QAOA-1.

In conclusion, we wish to underline that --- in light of the discussion upon the transferability of the Ansatz (see Fig.~\ref{fig:smooth_protocols}) --- the specific procedure adopted to obtain QAOA-2 solutions becomes less relevant: once a detailed study of a single sample is carried out, its optimized smooth solutions serve as an excellent {\em Ansatz}  for all other randomly generated training sets, yielding an effective unique procedure to find smooth QAOA solutions outperforming optimal digitized-QA, valid for both $\ncost=0,1$.

Concerning our study on randomized samples, we proceeded with the same iterative interpolation strategy starting from $P=16$, for both $\ncost=0,1$.
Once a smooth QAOA-2 solution is obtained for the first sample, the transferability of the {\em Ansatz}  yields a well-defined strategy to apply QAOA on all the other randomized samples.

\begin{figure}[ht]
\centering
\includegraphics[width=0.485\textwidth]{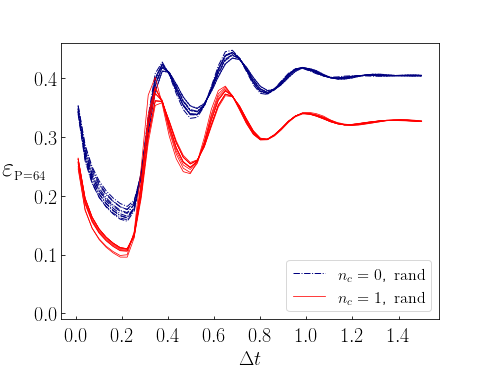}
\caption{
The one-dimensional landscape, versus $\Delta t$, of the variational energy density $\varepsilon_{\Ptrot}(\bbeta,\bgamma)$, Eq.~\eqref{eq:epsilon_P}, for the digitized-QA parameters, Eq.~\eqref{eq:dQA_angles}.  
All \emph{randomized} samples and both choices of $\ncost=0,1$ are shown. We remark that the landscape, and in particular the position of global minima, show mild sample-to-sample variability.     
}
\label{fig:dQA_summary_rand}
\end{figure}

\begin{figure*}[ht]
\centering
\includegraphics[width=0.485\textwidth]{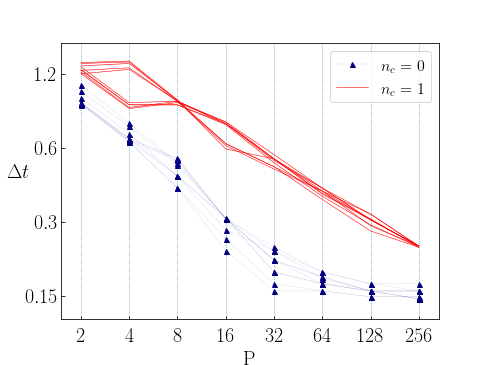}
\includegraphics[width=0.485\textwidth]{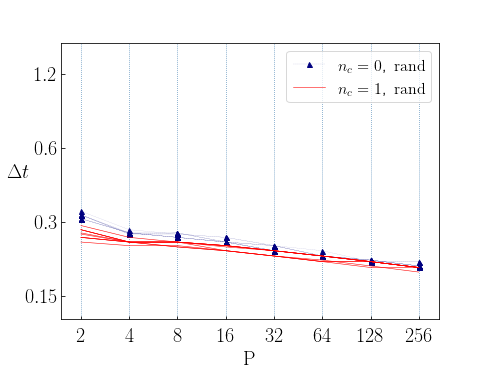}
\caption{
Optimal digitized-QA $\Delta t$ values for increasing $\Ptrot$, with logarithmic scales on both axes. The original samples case (left) compared to the randomized samples (right). The presence of evident clusters shows mild sample-to-sample variability, with few exceptions (as $\Ptrot=4$ for $\ncost=1$, in the left panel) due to an almost-flat energy landscape in that range.}
\label{fig:optimal_dQA_vsP}
\end{figure*}

\section{Additional results on digitized-QA} \label{app:randomized_dQA}

In this section, we report additional numerical results on optimal-$\Delta t$ digitized-QA, in particular by drawing a comparison between ordered and randomized samples.

In Fig.~\ref{fig:dQA_summary_rand} we show that, also in the randomized scenario, the $\Delta t$-landscape and the position of minima are almost identical for all samples in exam: we show data for $\Ptrot=64$ and both definitions of the cost-function $\ncost=0,1$, to be compared with Fig.~\ref{fig:dQA_summary} for the original samples.

The validity of these results naturally extends to different values of $\Ptrot$, as summarized in Fig~\ref{fig:optimal_dQA_vsP}, where we display the optimal values of $\Delta t$ vs $\Ptrot$ for the original samples (left panel) as well as for randomized samples (right panel).
In the latter case, we notice that the sample-to-sample variability of the optimal values of $\Delta t$ is even smaller, and also the differences --- for any fixed value of $\Ptrot$ --- between $\ncost=0,1$ are negligible (especially for large values of $\Ptrot$).
Apparently, by scrambling the landscape geometry, the initial specification of the cost-function becomes less relevant.
In contrast, we remark that the optimal values of $\Delta t$ differ significantly between any original sample and its randomized version.


\newpage

\bibliography{BiblioQAOA, BiblioLRB, BiblioQA_ter, BiblioQIC, BiblioQIsing, BiblioQSL, PTbib}


\end{document}